# Water-Induced Bimetallic Alloy Surface Segregation: A First Principle Study

Beien Zhu, Yi Gao*

*Division of Interfacial Water and Key Laboratory of Interfacial Physics and Technology, Shanghai Institute of Applied Physics, Chinese Academy of Sciences, Shanghai, 201800 China*

*Email: gaoyi@sinap.ac.cn

**Bimetallic alloys have drawn extensive attentions in materials science due to their widespread applications in electronics, engineering and catalysis. A very fundamental question of alloy is its surface segregation phenomenon. Many recent observations have shown that reactive gases or supports may have strong effects on alloy segregation. However, segregation in water—the most common solvent and environment—has not received enough attention. In this paper we give the quantitative descriptions on the surface segregation energies of 23 transition-metal impurities in Cu hosts under the conditions of water adsorption by performing density functional theory (DFT) calculations. The general trends in the changes of segregation energies caused by water adsorption are established. Our results show water adsorption could induce strong surface segregation tendencies for early and middle transition metals in Cu alloys. This finding not only prompts us to re-examine the potential effects of water on bimetallic alloy surfaces, but would be also very helpful as a guide for the further theoretical and experimental studies in this field.**

**PACS numbers: 64.75.Nx; 61.66.DK; 71.20.Be; 68.35.bd**

Metallic Alloy is one of the most common materials in our daily lives. In recent years, nanoalloys especially bimetallic nanoparticles draw remarkable interest due to their widespread potential for optical [1], magnetic [2], and catalytic [3] applications. The surface segregation properties of nanoalloys are of primary interest in this field since they are important in determining the alloys surface structures and compositions, which decides their chemical reactivity and especially catalytic activity eventually [4, 5]. Normally it can be predicted that the component with lower surface energy to be on the surface of one nanoalloy, however, this tendency can be changed by the external environment.

With the fast development of *in-situ* techniques in recent years [6-9], it has been proved that the reactive gas environment (i.e. CO, O, NO, etc.) and metal-support interaction can induce the surface segregation of the more active alloy component [9, 10-17]. The experiments on RhPd [6], AuPd [16] and CoPt [9, 12, 14] alloy catalysts have shown the changed segregation trends lead to unique surface structures and compositions under working conditions. The surface reconstruction is so dramatic that changes their physical and chemical properties completely. A "material gap" was found that the changes occurring in real catalysts cannot be truly represented by the model systems [10]. Despite of the significant progress achieved in this field [18-21], many important issues are open. In particular, water, the ubiquitous solvent and environment in reactions, has not received enough attention yet.

It has been well-known that the water-oil interface plays an important role in some organic reactions, referred as the "on water" catalytic effect by Sharpless and Marcus [22, 23]. Thus, is it possible for water to have a direct effect on the solid surface? Until now, there are very limited studies on this topic. Hansen and his co-workers reported the remarkable effect of water vapor on the morphology changes of supported copper nanocrystals [24]. Valden et al. observed a possible $H_2O$ induced Cr segregation in CrFe alloy [25]. Artrith and Kolpak used a highly accurate neural network potentials to study the equilibrium surface structure and composition of AuCu nanoalloys in aqueous solvents [26]. These recent studies show the potential effect of water on solid surface, however, two questions are still unclear. What is the effect of water on bimetallic alloy surface segregation trends? On which alloy combination the effect would be strong? It is the purpose of this work to answer these two questions by giving the quantitative description on the segregation energy under the condition of water adsorption.

In this work, we will show that the adsorption of water could alter the surface segregation properties of Cu-based alloys by first-principle calculations. The 23 transition-metal (TM) elements (the 3d (Ti-Ni), 4d (Zr-Ag), and 5d (Hf-Au) elements) are considered as the impurities in Cu substrates. Cu (111) surface and Cu (110) surface are chosen as the models since they are of great interest in studying metal-water interaction [27]. A database of the segregation energy in the dilute limit has been established by systemic calculations. Such a database can be a useful guide for the further theoretical and experimental studies in this aspect. Some general trends in the effect of water adsorption on the surface segregation energy have been revealed. In particular, a statistical model was applied to the MoCu alloy to confirm our prediction based on the thermodynamic framework.

The Vienna Ab initio Simulation Package (VASP) has been used to perform the DFT calculations [28]. Spin-unrestricted density functional theory (DFT) in the generalized gradient approximation (GGA) is employed [29]. The projector augmented-wave method (PAW) is used to describe the interactions between the valence electrons and the ionic cores [30, 31]. The plane-wave expansion is converged with a cut-off of 400 eV. The convergence for the electronic self-consistent is set to $10^{-5}$ eV. Geometry optimizations are performed within a conjugate-gradient algorithm with a convergence criterion on forces ($10^{-2}$ eV/Å). To model the (111) and (110) surfaces the periodic 2-D slab supercell approach is adopted. The slabs consist of 8 atomic layers and 4x4 unit cells separated by 15 Å of vacuum space for both surfaces. All metallic atoms in the top six layers of the slab and the water molecule/cluster are fully relaxed while the bottom two layers are constrained at the bulk geometry. The lattice parameter is set as 3.615 Å for Cu and 3.524 Å for Ni [32]. The Brillouin zone integrations for flat surfaces are performed on a Monkhorst-Pack (2x2x1) k-point mesh.

The surface segregation energy in the dilute limit, $E_{seg}$, is defined as the energy difference of moving the single impurity from bulk to surface, which has been proved a typical value to estimate the general trends in the surface segregation phenomena in TM alloys [4]. In this work it is calculated as below:

$$E_{seg} = E_{surf} - E_{4th-layer} \qquad (1),$$

where $E_{surf/4th-layer}$ corresponds to the energy that one site of the surface/4th-layer is the impurity while all the other sites are Cu. $E_{4th-layer}$ is considered as $E_{bulk}$ since the energy converges from $4^{th}$ layer, seen in Table S1. In the following discussion the superscript $v$ and $w$ represent for vacuum and for water adsorption respectively. This model has been adopted successfully to study the segregation phenomena of Au based alloys [33-35].

Our calculated vacuum $E_{seg}$ on Cu (111) surface are presented by the black squares in Fig 1(a). Consisting with the previous study [4], our results show the surface antisegregation tendencies ($E_{seg} > 0$) for most of the impurities on Cu (111) surface except for Zr, Pd, Pt, Ag and Au. It is reasonable since Cu, as a late transition metal, has lower surface energy than most of the transition metals [36, 37]. On the other hand, the surface segregation tendencies of Au and Ag in Cu can be predicted from our calculations, which agree with the experimental observations of $Cu_{core}$-$Ag_{shell}$ and $Cu_{core}$-$Au_{shell}$ nano-clusters [38, 39]. A quantitative comparison between our results and the data from Ref. 4 is shown in Fig. S1. No significant difference is found.

The proper adsorption sites of water monomer are tested, as shown in Fig. S2 (a). The top site, bridge site and 3-fold sites (hcp and fcc sites) of the Cu (111) surface and Cu alloy (111) surface were examined. In all the cases, the top sites are the most favorable adsorption sites, which are stronger in $H_2O$ adsorption than other sites about 0.1 eV for the Cu surface and even are the only stable adsorption sites for the alloy surfaces. These results are in good agreement with previous calculations on $H_2O$-metal interaction [40, 41]. Besides, the adsorption energies are not sensitive to the orientation of the water molecule (Fig. S2. (b)). Thus, the top site is chosen as the adsorption site in the following calculations and the H₂O molecule is laid parallel to the surface. In order to measure the effect of water adsorption on surface segregation, the H₂O molecule is always bound to the impurities when they are exchanged to the surface.

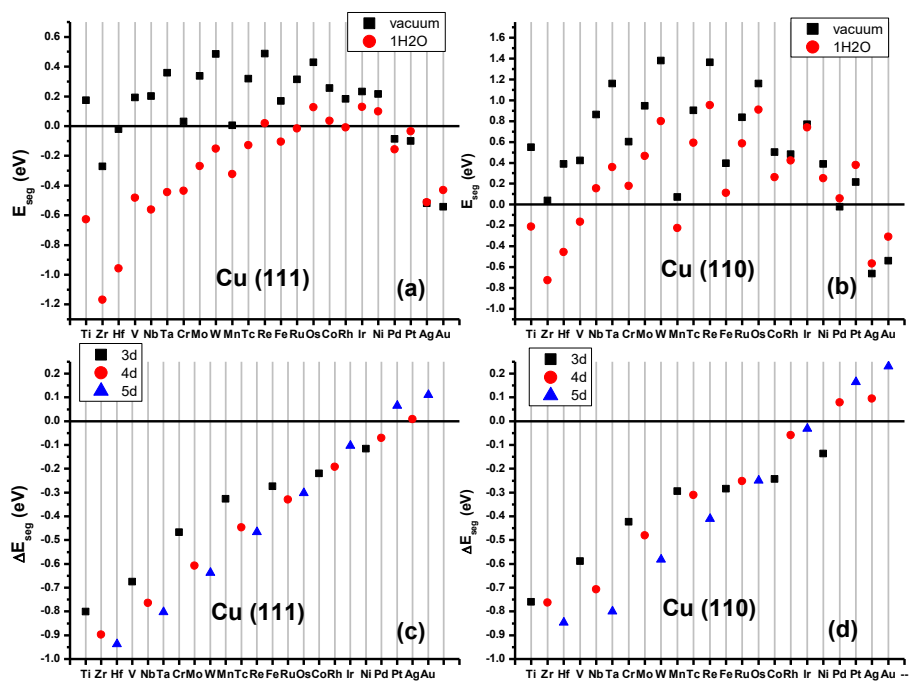

**Fig. 1.** Surface segregation energies of transition-metal impurities on (a) Cu (111) surface, (b) Cu (110) surface under vacuum (black squares) and with water monomer adsorption (red circles). The corresponding segregation energy differences caused by H₂O adsorption are shown in (c) and (d), respectively. The transition metals are sequenced according to their orders in the periodic table.

Interestingly, the $E_{seg}$ of different TMs on Cu (111) surface are changed to different degrees by water monomer adsorption. As shown in Fig. 1(a), with water adsorption a clear decrease of $E_{seg}$ can be found in most cases except those of Pd, Pt, Ag and Au. In particular, the values of $E_{seg}$ change from positive to negative for most early and middle transition metals, which indicates the surface segregation tendencies of these metals on Cu (111) surface induced by water monomer adsorption. The effects of water monomer adsorption are smaller for later transition metals. For Au, it can be seen that its surface segregation tendency is reduced by water adsorption. This is consistent with Artrith's recent study that mixed Au-Cu surfaces are thermodynamic preferred in aqueous solution while $Cu_{core}$-$Au_{shell}$ nanoparticles are favored under vacuum [26]. It should be noted that $E_{seg}$ are insensitive to the dispersion corrections, as shown in Fig. S3.

Besides the Cu (111) surface, we find the similar trends in changing the surface segregation energies by water monomer adsorption on Cu (110) surface as well (Fig. 1(b)). The three early TM impurities (Ti, Zr, Hf) and one middle TM impurity (V) show reversed surface segregation tendency under water monomer adsorption. The rest of the middle TM impurities, however, still prefer to be in the bulk of (110) surface. This is interesting because these impurities show opposite segregation tendencies on Cu (111) surface and on (110) surface under water monomer adsorption, while in vacuum they show antisegregation tendencies on both surfaces. This indicates the potential possibility to manipulate the concentrations of different facets of one Cu alloy nanoparticle with water adsorption, for example, a CuMo nanoparticle with the concentrated Mo on (111) facet and concentrated Cu on (110). It could be very useful in selecting catalysis and some other industrial applications.

To further investigate the trends in changing the $E_{seg}$ by water adsorption, we calculate the difference of the segregation energy ($\Delta E_{seg}$) defined as:

$$\Delta E_{seg} = E_{seg}^w - E_{seg}^v \qquad (2).$$

As shown in Fig. 1(c) and 1(d), the $\Delta E_{seg}$ on both Cu (111) surface and Cu (110) surface show regular linear trends with the orders of the impurities in the periodic table. The effect of water monomer adsorption increases linearly when the impurity moves to the left from silver and gold in the periodic table.

Concerning $E_{seg}^w$ can be described as:

$$E_{seg}^w = E_{seg}^v + E_{ads}^{impurity} - E_{ads}^{host} \qquad (3),$$

thus,

$$\Delta E_{seg} = E_{ads}^{impurity} - E_{ads}^{host} = \Delta E_{ads} \qquad (4),$$

where $\Delta E_{ads}$ represents for the adsorption energy difference between $H_2O$ on the surface impurity ($E_{ads}^{impurity}$) and on the host surface with a bulk impurity ($E_{ads}^{host}$). Therefore, the linear trends actually refer to the bonding strength of water monomer to different metals since $E_{ads}^{host}$ ($E_{ads}^{Cu}$ in this case) is a constant. According to the *d*-band model of Nøskov and Hammer, the trends in molecular chemisorption bonds to different metals can be described by the metals' *d*-band centers [42]. Thus the $\Delta E_{seg}$ in (111) surface is written as a function of the *d*-band center in Fig. 2 and the linear trends are found expectedly. As we move the impurities to the left from copper, silver or gold in the periodic table, the *d*-bands move up in energy and the bonds between the water monomer and the metals become stronger. Since the shift of the *d*-band center from one metal to another is one of the intrinsic chemical properties of transition metals, the linear trends of $\Delta E_{seg}$ should be independent on the facet, which explains the similar linear trends found both in Cu (111) surface and (110) surface. Moreover they should not be limited in Cu alloys either. When we change to other metallic substrate different from Cu, the linear trends shall shift in energy and the offset, $E_{ads}^{host}$, changes. This conclusion is confirmed by the $E_{seg}$ of 4d TM impurities in Ni (111) surface, as shown in Fig. S4. Similarly, the surface segregation energies of these Ni-based alloys are changed by water adsorption as well. More importantly, the linear trends of $\Delta E_{seg}$ are observed as expected.

From the linear trends a general conclusion can be made that if the host and the impurity are nearby in the periodic table, indicating close *d*-band centers, the $\Delta E_{seg}$ shall be small, otherwise it shall be large. Most importantly, the linear trends show the $\Delta E_{seg}$ could reach such a strong degree that the effect of water adsorption on the segregation properties of bimetallic alloys should be considered seriously in many related fields.

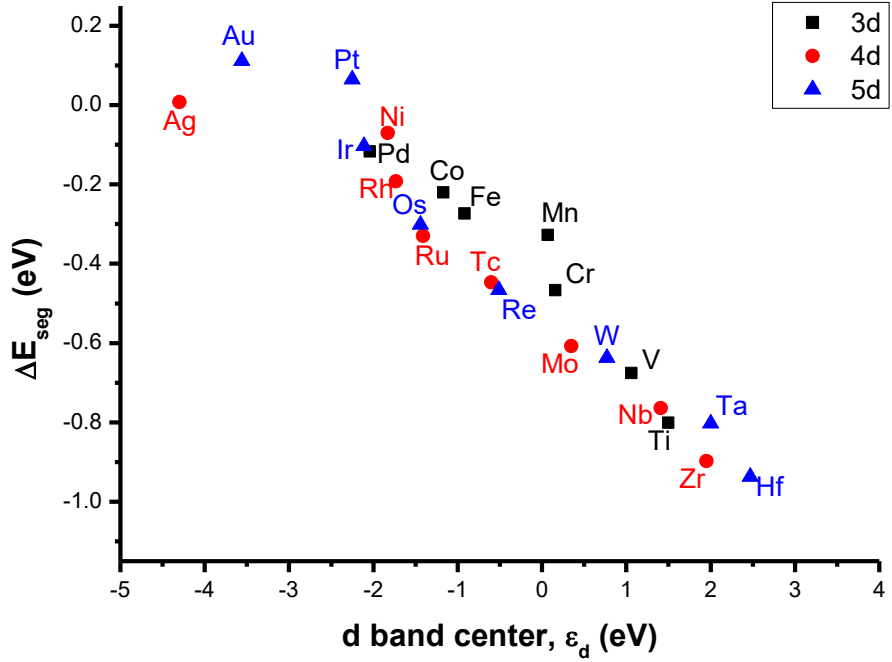

**Fig. 2.** Segregation energy differences of different transition metals in Cu (111) surface as a function of the metals' *d*-band centers (refer to the Fermi level).

In order to mimic the water vapor and humid environment, the $E_{seg}$ under the adsorptions of trimer and hexamer $H_2O$ clusters are calculated for all 23 Cu alloys. Considering the slabs we used, the trimer and hexamer water clusters refer to around 0.28 and 0.56 monolayer (ML) coverage, respectively. The stable structures of these water clusters on Cu (111) surface are gained from A. Michaelides's work [43]. To investigate the ideally maximum effect of water on alloy surface segregation, the surface impurity is bound to one $H_2O$ molecule of the clusters as shown in Fig. S5.

The results in Fig. 3 show the adsorption of water clusters can alter the segregation trends of transition-metals in Cu as well as water monomer. From water monomer to trimer, the changes of $E_{seg}$ are mostly smaller than 0.1 eV, except for the cases of Zr, Ta and W. In these cases much stronger water-induced surface segregation tendencies are found. The phenomena are more obvious for water hexamers. All of the early and middle transition metal impurities show particularly strong surface segregation tendencies under the adsorptions of hexamers. Such strong tendencies allow us to predict the migration of earlier transition-metals to the surface of Cu nanoalloys in a humid environment with high possibility.

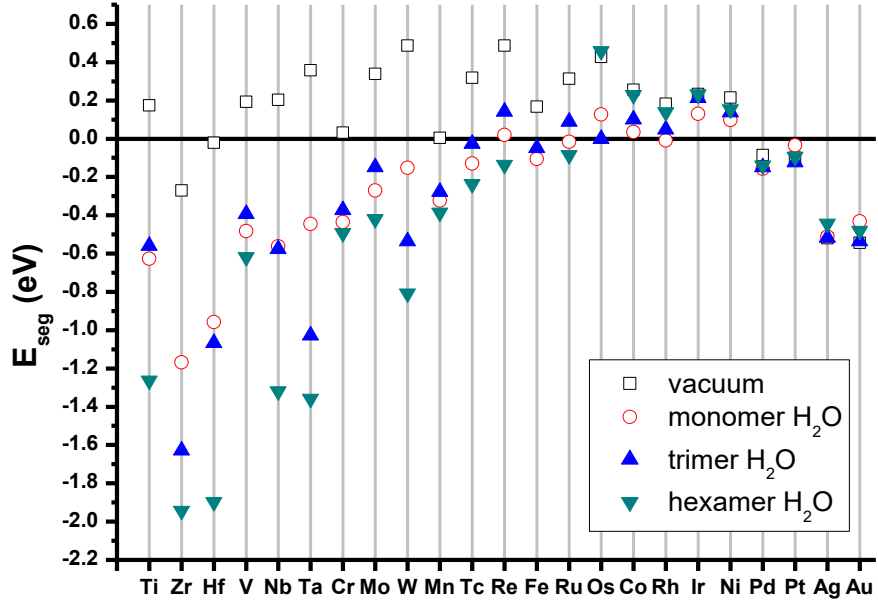

**Fig. 3**, surface segregation energies of transition metals in Cu (111) surface under vacuum and under the adsorption of water monomer, trimer and hexamer.

A statistical model is established to sample the Mo distribution in a MoCu icosahedra nanoparticle (NP) with certain water coverage and at certain temperature (*T*). The Metropolis method is employed to take into account the effect of configurational entropy [44]. The nanoparticle contains 411 Mo atoms and 1646 Cu atoms, 2057 atoms in total (Mo:Cu = 1:4). Mo is chosen as an example because from our data it shows reversed segregation tendencies on Cu (111) surface with water adsorption. An icosahedra NP is used as a model since all of its facets are (111) oriented. This particle is divided into four regions, the surface layer (R1), the sub-surface layer (R2), the sub-sub-surface layer (R3) and the core (R4). With this model we did this kind of trial: exchanging the chemical type of one randomly chosen atom *i* in one region (i.e. R2) with another randomly chosen atom *j* in its nearest region (i.e. R3 or R1). If a Mo atom is exchanged from the *i*th region to the *j*th region, the exchange energy is calculated like this:

$$\Delta E_{ij} = E(R_j, iw) - E(R_i, iw) \tag{5},$$

where $E(R_i, iw)$ is the energy of Mo inserted in the *i*th layer of Cu (111) surface, *iw* is the ratio of adsorbed water molecules to surface sites, estimated from water coverage. According to the experimental and theoretical data, monolayer water on Cu

(111) surface covers about 67% surface sites [45]. For a given water surface coverage, an exchange has a $iw$ chance to be carried out under the condition of water adsorption. Moreover, for a 1/3 ML coverage, the energies with trimer adsorptions are adopted, for the cases of 2/3 ML and 1 ML coverage, the energies with hexamer adsorptions are adopted. The detailed information can be found in Table S2. Finally, the acceptance possibility of each trial is calculated by:

$$Ratio = \exp(-\frac{\Delta E_{ij}}{k_b T}) \tag{6},$$

$k_b$ is the Boltzmann constant. Thus, the configurational entropy is involved when assessing the Mo stability in each region.

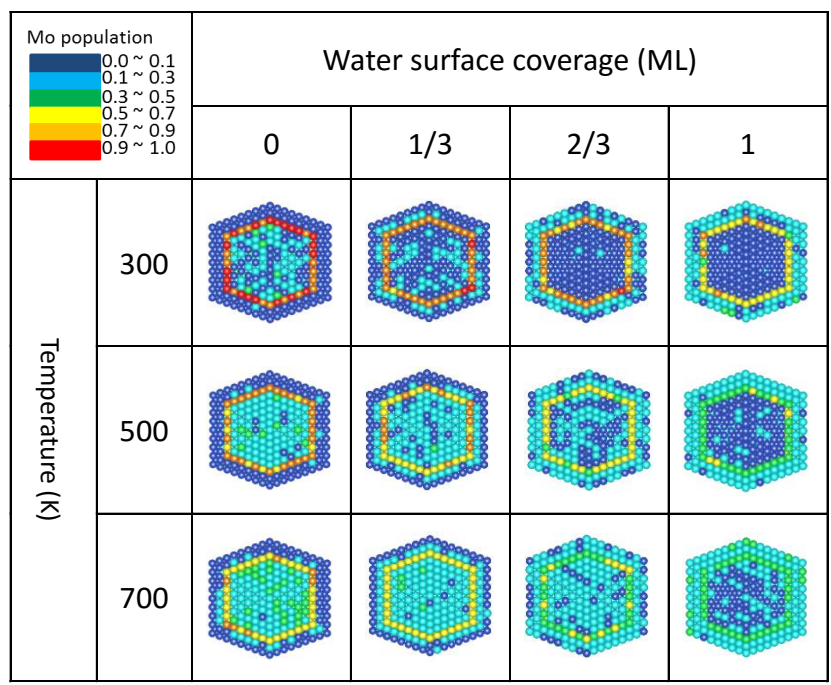

**Fig. 4,** cross sections of Icosahedra nano-particle containing 411 Mo atoms and 1646 Cu atoms (Mo:Cu = 1:4). The colour of each site is decided by the average Mo occupation rate, the colour map is inserted in the left top corner.

For each temperature and water coverage, we performed 1,000,000 trials for each region. The chemical type of each atomic site of the nanoparticle is counted every 100 times in the last 100,000 trials. Then the average Mo occupation rates of each site are calculated. It can be seen in Fig. 4 that without water adsorption Mo is blocked in the R3 due to its lower configuration energy. With water adsorption, Mo starts to migrate to the surface. The surface occupancy rate of Mo increases with the increasing temperature and surface water coverage, which can be seen more clearly with the more detailed data shown in Fig. S6. These statistical results confirm the prediction from the segregation energy calculations and indicate mixed Cu-Mo surfaces are thermodynamically favored in aqueous environment.

In summary, we report the changes in surface segregation energies of 23 transition-metals in two Cu surfaces caused by the adsorption of water monomer. For Cu alloys, the changes increase to a considerable degree when the impurities move to the left from Cu, Ag and Au in the periodic table. A general conclusion is made that the farther the two components of the alloy are in the periodic table, the stronger the effect of water adsorption on its surface segregation property. Further calculations with trimer and hexamer $H_2O$ clusters lead to the prediction that earlier transition metals prefer to be on the surfaces of Cu based alloys in a humid environment. This prediction is confirmed by a statistical model which indicates the thermo-stability of Mo atoms on the surface of CuMo nanoparticle with increasing water coverage and temperature. We hope this study would draw people's attention on the possible changed property of alloys in aqueous or humid environment and work as a foundation for the further studies in this field.

**ACKNOWLEDGEMENTS**

We thank for the discussion with Prof. Jun Hu and Prof. Hai-Ping Fang. BZ thanks for the development fund for Shanghai talents (Y439011011). YG thanks for the funding support from Shanghai Institute of Applied Physics, Chinese Academy of Sciences (Y290011011), National Natural Science Foundation of China (21273268), "Hundred People Project" from Chinese Academy of Sciences, and "Pu-jiang Rencai Project" from Science and Technology Commission of Shanghai Municipality (13PJ1410400). The computational resources utilized in this research were provided by Shanghai Supercomputer Center, National Supercomputing Center in Tianjin and National Supercomputing Center in Shenzhen.

**References：**

1. P. Mulvaney, Langmuir **12**, 788 (1996).

2. D. Alloyeau, C. Ricolleau, C. Mottet, T. Oitkawa, C. Langlois, T. Le Bouar, N. Braidy, A. Loiseau, Nat. Mater. **8**, 940 (2009)

3. P. Strasser, S. Koh, T. Anniyev, J. Greeley, K. More, Ch. Yu, Z. Liu, S. Kaya, D. Nordlund, H. Ogasawara, M. F. Toney, A. Nillson, Nat. Chem. **2**, 454 (2010).

4. A. V. Ruban, H. L. Skriver, J. K. Norskov, Phys. Rev. B 59, 15990 (1999)

5. G. Bozzolo, J. Ferrante, R. D. Noebe, B. Good, F. S. Honecy, P. Abel, Comput. Mater. Sci 15, 169 (1999)

6. F. Tao, M. E. Grass, Y. Zhang, D.R. Butcher, J.R. Renzas, Z. Liu, J.Y. Chung, B.S Mun, M. Salmeron, G.A. Somorjai, Science **322**, 932 (2008).

7. H.-G. Liao, L. Gui, S. Whitelam, H. Zheng, Science **336**, 1011 (2012).

8. H. Yoshida, Y. Kuwauchi, J. R. Jinschek, K. Sun, S. Tanaka, M. Kohyama, S. Shimada, M. Haruta, S. Takeda, Science **335**, 317 (2012).

9. H. L. Xin, S. Alayoglu, R. Tao, A. Genc, C.-M. Wang, L. Kovarik, E. A. Stach, L.-W. Wang, M. Salmeron, G. A. Somorjai, and H. Zheng, Nano Lett., **14**, 3203 (2014)

10. N. J. Divins, I. Angurell, C. Escudero, V. Pérez-Dieste and J. Llorca, Science **346**, 620 (2014)

11. V. Petkov, L. Yang, J. Yin, R. Loukrakpam, S. Shan, B. Wanjala, J. Luo, K. W. Chapman, C. J. Zhong, Phys. Rev. Lett., **109**, 125504 (2012)

12. F. Zheng, S. Alayoglu, J. Guo, V. Pushkarev, Y. Li, P.-A. Glans, J.-L. Chen, G. Somorjai, Nano Lett. **11**, 847 (2011).

13. M. A. Newton, C. Belver-Coldeira, A. Martinez-Arias, M. Fernandez-Garcia, Nat. Mater. **6 (7)**, 528 (2007).

14. D. Wang, H. L. Xin, R. Hovden, H. Wang, Y. Yu, D. A. Muller, F. J. DiSalvo, H. D. Abruña, Nat. Mater. **12 (1)**, 81 (2013).

15. H. L. Xin, J. A. Mundy, Z. Liu, R. Cabezas, R. Hovden, L. F. Kourkoutis, J. Zhang, N. P. Subramanian, R. Makharia, F. T. Wagner, D. A. Muller, Nano Lett. **12**, 490 (2011).

16. L. Delannoy, S. Giorgio, J. G. Mattei, C. R. Henry, N. Kolli, C. Methivier, C. Louis, ChemCatChem. **5**, 2707 (2013).

17. Y. T. Law, T. Skála, I. Píš, V. Nehasil, M. Vondráček, S. Zaferatos, J. Phys. Chem. C **116**, 10048 (2012)

18. F. Tao, M. Salmeron, Science **331**, 171 (2011).

19. S. Zafeiratos, S. Piccinin and D. Teschner, Catal. Sci. Technol. **2**, 1787 (2012).

20. S. Zhang, L. Nguyen, Y. Zhu, S. Zhan, C-K Tsung, F. Tao, Acc. Chem. Res. **46** 1731 (2013).

21. H. Guesmi, Gold Bull. **46**, 213 (2013).

22. S. Narayan, J. Muldoon, M. G. Ginn, V. V. Fokin, H. C. Kolb, K. B. Sharpless, Angew. Chem,. Int. Ed. **44**, 3275 (2005).

23. Y. Jung, R. A. Marcus, J. Am. Chem. Soc. **129**, 5492 (2007).

24. P. L. Hansen, J. B. Wagner, S. Helveg, J. R. Rostrup-Nielsen, B. S. Clausen, H. Topsøe, Science **295**, 2053 (2002).

25. P. Jussila, H. Ali-Löytty, K. Lahtonen, M. Hirsimäki and M. Valden, Surf. Sci. **603**, 3005 (2009).

26. N. Artrith, A. M. Kolpak, Nano. Lett. **14**, 2670 (2014).

27. J. Carrasco, A. Hodgson, A. Michaelides, Nat. Mater. 11, 667 (2012).

28. G. Kresse, J. Hafner, Phys. Rev. B 47, 558 (1993).

29. J. P. Perdew, K. Burke, M. Ernzerhof, Phys. Rev. Lett. 77, 3865 (1996).

30. P. E. Blochl, O. Jepsen, O. K. Andersen, Phys. Rev. B 49, 16223 (1994).

31. G. Kresse, D. Joubert, Phys. Rev. B 59, 1758 (1999).

32. C. Kittel, "Introduction to Solid State Physics", Wiley, New York (1971).

33. H. Guesmi, C. Louis, L. Delannoy, Chem. Phys. Lett., **503** (2011) 97.

34. A. Dhouib, H. Guesmi, Chem. Phys. Lett., **521** (2012) 98.

35. B. Zhu, G. Thrimurthu, L. Delannoy, C. Louis, C. Mottet, J. Creuze, B. Legrand, H. Guesmi, Journal of Catalysis **308**, 272 (2013).

36. J. Friedel, Ann. Phys. (N.Y.) **1**, 257 (1976)

37. F. Cyrot-Lackmann, Adv. Phys. **16**, 393 (1967).

38. M. Cazayous, C. Langlois, T. Oikawa, C. Ricolleau, A. Sacuto, Phys. Rev. B 73,

113402 (2006).

39. L. M. Wang, S. Bulusu, H.-J. Zhai, X. C. Zeng, L. S. Wang, Angew. Chem., Int. Ed. 46, 2915 (2007).

40. J. Carrasco, B. Santra, J. Klimes, A. Michaelides, Phys. Rev. Lett., **106**, 026101 (2011).

41. J. Carrasco, J. Klimes and A. Michaelides, J. Chem. Phys. **138**, 024708 (2013).

42. B. Hammer and J. K. Norskov, Adv. Catal. **45**, 71 (2000).

43. A. Michaelides, Faraday Discuss. 136, 287 (2007).

44. N. Metroplis, A. W. Rosenbluth, M. N. Rosenbluth, A. H. Teller, E. Teller, J. Chem. Phys. 21, 1087 (1953).

45. A. Hodgson, S. Haq, Surf. Sci. Rep. 64, 381 (2009).